\documentclass[a4paper,final]{aipproc}
\layoutstyle{6x9}
\title{A model of the internal structure of Titan:first results}
\author{G.Pavi \v ci\'c} {address=
{Public Observatory,Gornji grad 16,11000 Beograd,Serbia and Montenegro}\\
$moskito752000@yahoo.com$}
\author{V. \v Celebonovi\'c} {address=
{Inst.of Physics,Pregrevica 118,11080 Zemun-Beograd,Serbia and Montenegro}\\
$vladan@phy.bg.ac.yu$}
\begin{document}
\begin{abstract}
The aim of this short contribution is to report some preliminary results on a theoretical model of the internal structure of Titan,the satellite of Saturn.All the calculations were performed within the $SK$ theory of dense matter.The central pressure and temperature,as well as a possible chemical composition of this satellite have been determined,but they will ultimately become verifiable only by data from the {\it Cassini} mission.
\end{abstract}
\date{\today}
\maketitle{}
\section{Introduction}
 Astronomers,physicist and even biologists have been interested in Titan for many years.The main motivation for this interest is the fact that Titan is the only satellite in the Solar System which has an atmosphere. It is widely thought in the research community that this atmosphere resembles the early atmosphere of the Earth,perhaps in pre-biotic times \cite{DEM:04},\cite{ROE:03} and earlier references given there. This continious interest was the main "`driving force"' for the preparation and launch of the {\it Cassini} mission.Titan's atmosphere is dense and opaque,so there exists no way to gain any information whatsoever about its surface from Earth based observation.Accordinly,the logical question which emerges concerns the possibility of obtaining some information about the surface and interior of Titan theoretically. 
 
The aim of this note is to briefly propose a theoretical model of the interior of Titan.It was calculated within the so-called $SK$ theory of dense matter,proposed by P.Savi\'c and Radivoje Ka\v sanin, \cite{SK:6265},and reviewed elsewhere in these proceedings \cite{CHE:04}. The following section contains a brief remainder of the main ideas of the $SK$ theory and the finl one is devoted to results of its applications to Titan.
\section{The SK theory:a quick introduction}
The development of what later became the SK theory started in 1961
when P.Savi\'c  published a short paper \cite{SAV:61} presenting an
unusual idea: the mean planetary densities,as calculated from the
masses and radii known at the time,could be linked to the mean solar
density by an extremely simple expression:
\begin{equation}
    \rho=\rho_{0}2^{\phi}
\end{equation}

In this expression $\rho_{0}$ denotes the mean solar density,which
at the time was estimated at $4/3$ g$cm^{-3}$. By choosing integral
values of the exponent $\phi$ in the interval $\phi\in(-2,2)$
Savi\'c managed to fit the numerical values of the densities of the
major planets. No hint of any possible physical explanation of this
simple fit was given. In the following 4 years,collaborating with
Radivoje Ka\v sanin, \cite{SK:6265} he managed to develop a theory
of the behavior of materials under high pressure.In later years,it
was nicknamed "the SK theory" after the first letters in their
family names.

The basic physical idea of their theory is simple.SK assumed that
sufficiently high values of external pressure lead to changes of the
electronic structure of atoms and/or molecules.In the course of time,two kinds of applications of this theory were developed.One concerns modellisation of celestial bodies,and the other is oriented towards the analysis of laboratory data on the behaviour of materials under high static pressure.

A final detail: in astronomical applications,this theory needs only a couple of input data: the mass and the radius of the object under study.

\section{Application to Titan}

The mass and radius of Titan were taken from the web site of the US National Space Science Data Center,at the address http://nssdc.gsfc.nasa.gov/. After the application of the algortighm of the theory,the following values of various parameters of Titan were obtained:

\begin{center}
$82 \leq A \leq86$
\end{center}

\begin{center}
$T^{*}=300 K$
\end{center}

\begin{center}
$p^{*}=0.01 MBar$
\end{center}

The central pressure and temperature are denoted by $p^{*},T^{*}$. The symbol $A$ denotes the mean molecular mass of the mixture of materials that the object under study is made of. It can be calculated within the $SK$ theory,but representing it with any combination of real chemical elements and/or compounds is a difficult task,which has (whenever possible) to be constrained by observed data.As in the case of Titan there exist no such data,the value of $A$ obtained  was represented by a combination of materials which seemed reliable,and which is

\begin{center}
$Fe_{2}SiO_{3}+FeS+SO_{2}+CH_{4}+N_{2}+Ir$
\end{center}

How close this combination is to "`real"' Titan will only become verifiable by forthcoming data from the {\it Cassini} mission.At the time of writing,the probe is orbiting Saturn,and it has alreday discovered two new moons.According to predictions,the first close fluby of Titan will occur towards the end of October of 2004.

\section{Acknowledgements}
The authors are grateful to Prof.Wim van Saarloos,director of the Lorentz Center for the initial proposal on the organization of this workshop,and to all in the Lorentz Center whose work helped that the workshop runs so smoothly and succesively.One of the authors (V.\v C) was financed by the Ministry of Science and Protction of the Environment of Serbia under its project 1231.

.

\end{document}